\documentclass[journal]{IEEEtran}

\usepackage{cite}

\usepackage{makeidx,epsfig}
\usepackage{setspace,graphicx,multirow}

\usepackage{amsfonts,latexsym,amssymb,amsthm}

\usepackage{graphicx}
\usepackage{epstopdf}

\usepackage{caption3} 
\DeclareCaptionOption{parskip}[]{} 
\usepackage[small]{caption}

\usepackage{subcaption}

\usepackage{array}

\usepackage[cmex10]{amsmath}
\DeclareMathOperator{\diag}{diag}  
\DeclareMathOperator{\var}{var}

\usepackage{url}

\usepackage{multirow}
\usepackage{hhline}

\usepackage{amsmath}
\usepackage{algorithm}
\usepackage[]{algpseudocode}
\usepackage{varwidth}

\makeatletter
\def\BState{\State\hskip-\ALG@thistlm}
\makeatother

\usepackage{algcompatible}

\usepackage{fixltx2e}

\newcommand*\diff{\mathop{}\!\mathrm{d}}

\usepackage{enumitem}

\makeatletter
  \newcommand\tinyv{\@setfontsize\tinyv{7pt}{9}}
\makeatother

\usepackage{authblk}

\usepackage{textcomp}

\usepackage{dsfont}

\usepackage{color}

\ifodd 0
\newcommand{\rev}[1]{{\color{red}#1}} 
\newcommand{\com}[1]{\textbf{\color{blue} (COMMENT: #1)}} 
\else
\newcommand{\rev}[1]{#1}

\newcommand{\com}[1]{}
\fi

\begin{document}
\bibliographystyle{IEEEtran}
\bstctlcite{IEEEexample:BSTcontrol}

\title{\LARGE Spatial Throughput Characterization for Intelligent Reflecting Surface Aided Multiuser System}

\author{Jiangbin~Lyu,~\textit{Member,~IEEE},
        and~Rui~Zhang,~\textit{Fellow,~IEEE}%
\thanks{This work was supported in part by the National Natural Science Foundation 
        	of China (No. 61801408 and No. 61771017), the Natural Science Foundation of Fujian 
        	Province (No. 2019J05002) and the Fundamental Research Funds for the Central 
        	Universities (No. 20720190008).}
\thanks{J. Lyu is with School of Informatics, and Key Laboratory of Underwater Acoustic Communication and Marine Information Technology, Xiamen University, China 361005 (e-mail: ljb@xmu.edu.cn); R. Zhang is with the Department of Electrical and Computer Engineering, National University of Singapore, Singapore 117583 (email: elezhang@nus.edu.sg).}
}

\markboth{IEEE Wireless Communications Letters,~Vol.~XX, No.~X, XXXXX~XXXX}%
{Lyu \MakeLowercase{\textit{et al.}}: Spatial Throughput Characterization for Intelligent Reflecting Surface Aided Multiuser System}


\maketitle

\begin{abstract}
Intelligent Reflecting Surface (IRS) has been recently proposed as a promising solution to enhance the spectral and energy efficiency of future wireless networks by tuning a massive number of low-cost passive reflecting elements and thereby constructing favorable wireless propagation environment.
Different from the prior works that focus on link-level performance optimization for IRS-aided wireless systems, this letter
characterizes the spatial throughput of a single-cell multiuser system aided by multiple IRSs that are randomly deployed in the cell.
It is shown by simulation that our analysis is valid and the IRS-aided system outperforms the full-duplex relay-aided counterpart system in terms of spatial throughput when the number of IRSs exceeds a certain value.
Moreover, it is shown that different deploying strategies for IRSs/active relays should be adopted for their respective throughput maximization.
Finally, it is revealed that given the total number of reflecting elements for IRSs, the system spatial throughput increases when fewer IRSs are deployed each with more reflecting elements, but at the cost of more spatially varying user rates.
\end{abstract}

\section{Introduction}

%
%
%
%
%
%

Intelligent Reflecting Surface (IRS) is a promising solution to achieve spectral and energy efficient, and yet cost-effective wireless networks in the future, by smartly reconfiguring the wireless propagation environment that is traditionally deemed to be random and uncontrollable\cite{QQirsMag},\cite{IRSbasar}.
By controlling the signal reflection via a massive number of low-cost passive elements, each being able to induce an amplitude and/or phase change to the incident signal independently, IRS can achieve various useful functions such as three-dimensional (3D) passive beamforming, spatial interference nulling and/or cancellation, and so on.
\rev{Compared to
the conventional active relaying/beamforming, IRS
requires much lower hardware cost and energy consumption
due to passive reflection, and yet operates spectral efficiently in full-duplex (FD)
without the need of costly self-interference cancellation (SIC)\cite{MarcoRelayMagazine}}.

The research on IRS is still at its early stage, but has recently attracted a great deal of attention.
The existing works on IRS-aided wireless systems mainly address challenges at the link level with given locations of one or more IRSs, which show that the IRS-aided system can achieve significant power saving or spectral efficiency improvement over the traditional system without IRS, by optimizing the IRS reflection coefficients \cite{QQtwc,IRSchauYuen,IRSjinShi,IRSqqDiscrete,IRSycLiang,IRSshuowen}.
\rev{In \cite{IRSemilDFrelay}, the authors compare the system performance aided by one single IRS or an active 
decode-and-forward (DF) relay deployed at the same location, which, however, ignores the fact that in practice the deployment strategy for active relay and passive IRS should be different.}
The authors in \cite{IRScuiYing} consider a multi-IRS-assisted system in which multiple IRSs at given locations adapt their phase shifts to minimize the user's outage probability.
\rev{In \cite{HanzoMIMOstochasticGeometry}, the authors investigate a single-cell multi-user system aided by multiple co-located intelligent surfaces (equivalent to a single large IRS).
However, the above works consider either a single IRS or multiple IRSs at given locations, and have not addressed the multi-IRS deployment issue, which is practically important for IRS-aided wireless systems\cite{QQirsMag}.}

\begin{figure}
\centering
   \includegraphics[width=0.95\linewidth,  trim=0 0 4 0,clip]{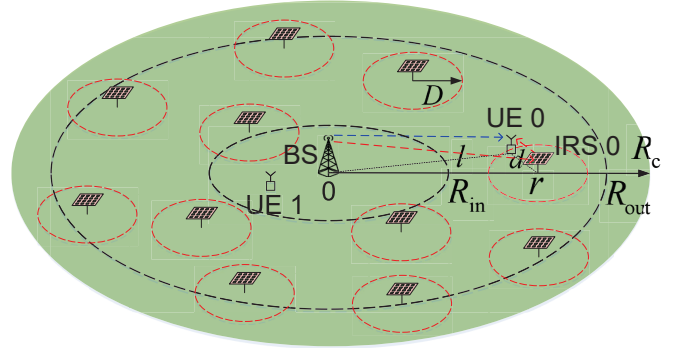}
\caption{\rev{IRS-aided multiuser communication in a single cell.\vspace{-2ex}}}\label{IRS}
\end{figure}

In this letter, we consider a single-cell system with one base station (BS) serving multiple user equipments (UEs), aided by multiple IRSs as shown in Fig. \ref{IRS}. We aim to characterize the \textit{spatial throughput} of the system, defined as the achievable rate per UE that has been averaged over the random IRS/UE locations as well as the wireless channel fading.
To this end, we first analyze the IRS-related fading channel statistics, and show that the IRS-reflected path exhibits the channel hardening effect and thus has an approximately constant power gain, denoted by $\kappa$, over the BS-UE direct path, which depends on the IRS-UE distance and the number of IRS elements.
Based on $\kappa$, we then propose a practical association rule for assigning UEs with their nearby IRS (if any).
Next, we derive a closed-form expression for the achievable rate of UEs averaged over channel fading, and then characterize the spatial throughput of the network by averaging over the random IRS and UE locations.
Finally, we evaluate the derived spatial throughput by simulations, and investigate the effect of key system parameters on the spatial throughput and the optimal IRS deployment strategy for maximizing it.
It is shown that the IRS-aided system outperforms a benchmark FD relay-aided system when the number of IRSs exceeds a certain value.
Different from active relays which are shown to be optimally deployed around the midpoint between the BS and their served UEs to balance the BS-relay and relay-UE links, IRSs mainly provide local coverage and should be deployed close to their served UEs. 
Moreover, it is shown that the optimal IRS deployment range from the BS depends on the number of IRSs/elements per IRS, as well as the UE signal-to-noise ratio (SNR) distribution in the cell.
Furthermore, it is shown that given the total number of reflecting elements for IRSs, it is beneficial to assemble more elements into fewer IRSs so as to maximize the system spatial throughput, but at the cost of more spatially varying rates of the UEs.

\section{System Model}\label{SectionModel}

Consider a single-cell multiuser system with one BS serving a group of $K$ UEs that are uniformly and randomly distributed in the disc cell area with radius $R_\textrm{c}$ meters (m) and centered at the BS, as shown in Fig. \ref{IRS}. 
We consider the downlink communication from the BS to UEs, whereas the results obtained can be similarly applied to the uplink communication as well. 
To facilitate our analysis, \rev{we assume that the transmission bandwidth and time for a given duration is equally divided into $K$ \textit{orthogonal} resource blocks (RBs), each assigned to one UE,} over which the channel is assumed to be flat-fading. 

A set of $M$ IRSs are deployed to assist the BS-UE communications in the cell, which are assumed to be uniformly and randomly deployed in the distance range $[R_\textrm{in},R_\textrm{out}]$ from the BS, where $0< R_\textrm{in}\leq R_\textrm{out}< R_\textrm{c}$ in practice.
\rev{Consider a typical UE 0 at any arbitrary location in the cell, whose nearest IRS is denoted by IRS 0, as shown in Fig. \ref{IRS}.
Denote $l$, $r$ and $d$ as the BS-UE 0, BS-IRS 0 and UE 0-IRS 0 horizontal distances, respectively.
Since IRS is effective in providing local signal enhancement, we propose a practical association rule that UE 0 is associated with its nearest IRS 0 if $d\leq D$, where the range $D$ will be specified later in Section \ref{SectionD}.}
To maximize the passive beamforming gain of the IRS to each served UE, we assume that its served UEs are assigned in orthogonal-time RBs, i.e., at each time instant, each IRS serves at most one UE.
On the other hand, those UEs outside range $D$ of any IRS is served by the BS only (e.g., UE 1).

\subsection{Channel Model}

Consider that the BS and UEs are each equipped with a single antenna, while each IRS has $N$ reflecting elements.
The baseband equivalent channels from the BS to IRS 0, from IRS 0 to UE 0, and from the BS to UE 0 are denoted by $\bold h_{\textrm{i}}\triangleq[h_{\textrm{i},1},\cdots,h_{\textrm{i},N}]^T\in \mathbb{C}^{N\times 1}$, $\bold h_{\textrm{r}}\triangleq[h_{\textrm{r},1},\cdots,h_{\textrm{r},N}]^T\in \mathbb{C}^{N\times 1}$, and $h_{\textrm{d}}\in \mathbb{C}$, respectively.
Let $\boldsymbol\phi\triangleq [\phi_1, \cdots,\phi_N]$ and furthermore denote $\boldsymbol\Phi\triangleq \diag\{ [e^{j\phi_1}, \cdots,e^{j\phi_N}]\}$ (with $j$ denoting the imaginary unit) as the phase-shifting matrix of IRS 0, where $\phi_n\in[0,2\pi)$ is the phase shift by element $n$ on the incident signal.\footnote{In this letter, we assume (maximum) unit amplitude for each reflection coefficient to maximize the IRS beamforming gain to its served UE\cite{QQtwc}.} The cascaded BS-IRS-UE channel is then modeled
as a concatenation of three components, namely, BS-IRS link, IRS reflecting with phase shifts,
and IRS-UE link, given by \cite{QQtwc}
\begin{equation}
h_{\textrm{ir}}\triangleq \bold h_{\textrm{i}}^T \boldsymbol\Phi \bold h_{\textrm{r}}.
\end{equation}

Assume that the cascaded channel phase $\angle(h_{\textrm{i},n}h_{\textrm{r},n})$ via each IRS element $n=1,\cdots,N$ can be obtained via channel estimation \cite{IRSzhengBeixiong}.
The IRS then adjusts the phase shift $\boldsymbol\phi$ such that the $N$ reflected signals are of the same phase at its served UE's receiver by setting $\phi_n=-\angle(h_{\textrm{i},n}h_{\textrm{r},n}), n=1,\cdots,N$.
As a result, we have
\begin{equation}
|h_{\textrm{ir}}|=|\bold h_{\textrm{i}}|^T |\bold h_{\textrm{r}}|=\sum\limits_{n=1}^N |h_{\textrm{i},n}| |h_{\textrm{r},n}|.
\end{equation}
We further assume that the BS-UE channel phase $\angle h_{\textrm{d}}$ is also known and the IRS can perform a common phase-shift such that $h_{\textrm{ir}}$ and $h_{\textrm{d}}$ are co-phased and hence coherently combined at the UE\cite{QQtwc}, with the overall channel amplitude denoted by $Z\triangleq |h_{\textrm{ir}}|+|h_{\textrm{d}}|$.

For the BS-IRS, IRS-UE and BS-UE links, we assume a block-fading channel which consists of distance-dependent path-loss with path-loss exponent $\alpha\geq 2$ and an additional random term $\xi$ accounting for small-scale fading.
The BS-UE channel power gain is thus given by
\begin{equation}\label{gau}
|h_{\textrm{d}}|^2\triangleq g_{\textrm{d}}\xi_{\textrm{d}}=\beta (l^2+H_\textrm{B}^2)^{-\alpha/2}\xi_{\textrm{d}},
\end{equation}
where $g_{\textrm{d}}$ is the average channel power gain, $H_\textrm{B}$ denotes the height of the BS and $\beta=(\frac{4\pi f_c}{c})^{-2}$ denotes the average channel power gain at a reference distance of 1 m, with $f_c$ denoting the carrier frequency, and $c$ denoting the speed of light.
$\xi_{\textrm{d}}\sim \textrm{Exp}(1)$ is an exponential random variable (RV) with unit mean accounting for the small-scale Rayleigh fading.\footnote{\rev{The proposed method can be extended to account for other fading channels.}}
Accordingly, $|h_{\textrm{d}}|$ follows the Rayleigh distribution with scale parameter $\sqrt{g_{\textrm{d}}/2}$, denoted by $\mathcal{R}\big(\sqrt{g_{\textrm{d}}/2}\big)$.

Similarly, the channel power gains from the BS to the $n$-th element of IRS 0, and from the IRS to UE 0 are given by
\begin{equation}\label{gai}
|h_{\textrm{i},n}|^2\triangleq g_{\textrm{i}}\xi_{\textrm{i},n}=\beta \big(r^2+(H_\textrm{B}-H_\textrm{I})^2\big)^{-\alpha/2}\xi_{\textrm{i},n},
\end{equation}
and
\begin{equation}\label{giu}
|h_{\textrm{r},n}|^2\triangleq g_{\textrm{r}}\xi_{\textrm{r},n}=\beta \big(d^2+H_\textrm{I}^2\big)^{-\alpha/2}\xi_{\textrm{r},n},
\end{equation}
where $g_{\textrm{i}}$ and $g_{\textrm{r}}$ denote the respective average channel power gains, and $H_\textrm{I}$ denotes the height of the IRS.\footnote{For simplicity, we assume $H_\textrm{B}\geq 1$ m and $H_\textrm{I}\geq 1$ m to avoid unbounded power gain when the horizontal distance $l$ or $d$ goes to zero.}
Therefore, we have $|h_{\textrm{i},n}|\sim\mathcal{R}\big(\sqrt{g_{\textrm{i}}/2}\big)$ and $|h_{\textrm{r},n}|\sim\mathcal{R}\big(\sqrt{g_{\textrm{r}}/2}\big)$.

\subsection{Achievable Rate and Spatial Throughput}

Denote the downlink transmit power for each UE as $P_0$.
If the typical UE 0 is served by the BS only, then its received SNR is given by 
\begin{equation}\label{gamma}
\gamma\triangleq |h_{\textrm{d}}|^2 P_0/\sigma^2=|h_{\textrm{d}}|^2\gamma_0,
\end{equation}
where $\gamma_0\triangleq P_0/\sigma^2$, and $\sigma^2$ denotes the receiver noise power.
On the other hand, if UE 0 is also served by IRS 0, we have 
\begin{equation}\label{gamma2}
\gamma\triangleq Z^2 \gamma_0=(|h_{\textrm{ir}}|+|h_{\textrm{d}}|)^2 \gamma_0.
\end{equation}
As a result, the achievable rate in bits/second/Hz (bps/Hz) averaged over the fading distribution is given by
\begin{equation}\label{Ck}
C\triangleq\mathbb{E}\{\log_2( 1+ \gamma)\}.
\end{equation}
Finally, denote $\bar C\triangleq \mathbb{E}\{C\}$ as the \textit{spatial throughput} averaged over the distributions of all UE and IRS random locations.

\section{Spatial Throughput Characterization}\label{SectionLink}
In this section, we first derive the channel statistics and IRS coverage/serving range $D$, based on which we then characterize the spatial throughput $\bar C$ of the IRS-aided multiuser system.
Last, we introduce a relay-aided benchmark system without using IRS.

\subsection{Channel Statistics}

\rev{Assume that the fading channels $h_{\textrm{d}}$, $h_{\textrm{i},n}$ and $h_{\textrm{r},n}$, $n=1,\cdots,N$
are independent.}
Then for the BS-IRS-UE signal that goes through element $n$, the channel amplitude is subject to double-Rayleigh fading given by
\begin{equation}
|h_{\textrm{ir},n}|\triangleq |h_{\textrm{i},n}| |h_{\textrm{r},n}|.
\end{equation}
Note that for a double-Rayleigh distributed RV $Y=X_1X_2$ with independent $X_1\sim \mathcal{R}(\delta_1)$ and $X_2\sim \mathcal{R}(\delta_2)$, its mean and variance are respectively given by
\begin{equation}
\mathbb{E}\{Y\}\triangleq \pi\delta_1\delta_2/2,
\end{equation}
\begin{equation}
\var\{Y\}\triangleq 4\delta_1^2\delta_2^2(1-\pi^2/16).
\end{equation}
Therefore, we have
\begin{equation}
\mathbb{E}\{|h_{\textrm{ir},n}|\}\triangleq\frac{\pi}{4}\sqrt{g_{\textrm{i}}g_{\textrm{r}}},
\end{equation}
\begin{equation}
\var\{|h_{\textrm{ir},n}|\}\triangleq(1-\pi^2/16)g_{\textrm{i}}g_{\textrm{r}}.
\end{equation}

\rev{Since the channel amplitudes $|h_{\textrm{ir},n}|$, $n=1,\cdots,N$ are independent and identically distributed (i.i.d.),
by the central limit theorem (CLT), the composite amplitude for the BS-IRS-UE channel for practically very large\footnote{\rev{We consider electrically small IRSs\cite{MarcoRelayMagazine}, where each reflecting element is typically bounded within a square region of side length around $1/5$ wavelength. Therefore, to fit into the size of 1 m$^2$ at $f_c=2$ GHz, we have $N>1000$, while it can be even larger at higher frequency.}} $N$ can be approximated by the Gaussian distribution}, i.e.,
\begin{align}
&|h_{\textrm{ir}}|=\sum_{n=1}^N |h_{\textrm{ir},n}|\stackrel{\textrm{approx.}}{\sim} \mathcal{N}\big(N\mathbb{E}\{|h_{\textrm{ir},n}|\},N\var\{|h_{\textrm{ir},n}|\}\big)\notag\\
&=\mathcal{N}\big(N\frac{\pi}{4}\sqrt{g_{\textrm{i}}g_{\textrm{r}}},\quad N(1-\pi^2/16)g_{\textrm{i}}g_{\textrm{r}}\big).\label{haiu}
\end{align}
Denote $\mu$ and $\omega$ as the mean and standard deviation of $|h_{\textrm{ir}}|$, respectively, whose ratio is given by
\begin{equation}\label{ratio}
c_0\triangleq\frac{\mu}{\omega}=\frac{\pi\sqrt{N}}{\sqrt{16-\pi^2}},
\end{equation}
which is a function of $N$ only. \eqref{ratio} implies a ``channel hardening" effect where the mean $\mu$ increases faster than the standard deviation $\omega$ as $N$ increases, thus resulting in relatively less variations around the mean value. \rev{Channel hardening corresponds to a nearly deterministic channel with improved reliability, and requiring less frequent channel estimation.}

As a result, the normalized channel power $|h_{\textrm{ir}}|^2/\omega^2$ follows the noncentral chi-square distribution with mean $1+(\mu/\omega)^2$. The average BS-IRS-UE channel power is thus given by
\begin{equation}\label{gir}
g_{\textrm{ir}}\triangleq \mathbb{E}\{|h_{\textrm{ir}}|^2\}=\omega^2+\mu^2= G_\textrm{bf} g_{\textrm{i}}g_{\textrm{r}},
\end{equation}
which is proportional to the average channel power product $g_{\textrm{i}}g_{\textrm{r}}$ with the beamforming gain factor $G_\textrm{bf}\triangleq \frac{\pi^2}{16}N^2+\big(1-\frac{\pi^2}{16}\big)N$. This is consistent with the $O(N^2)$ scaling law shown in \cite{QQtwc}.

Finally,
the composite channel amplitude $Z$ is the sum of a Gaussian RV and an independent Rayleigh RV, whose probability density function (pdf) can be obtained via the convolution of their individual pdfs. However, the resultant pdf of $Z$ does not render a closed-form integral value for the achievable rate $C$ in \eqref{Ck}.
To tackle this difficulty, we approximate the distribution of $Z^2$ by the Gamma distribution, whose mean and variance are respectively given by
\begin{equation}\small
\mathbb{E}\{Z^2\}\triangleq \mathbb{E}\{(|h_{\textrm{ir}}|+|h_{\textrm{d}}|)^2\}=G_\textrm{bf} g_{\textrm{i}}g_{\textrm{r}}+N\frac{\pi}{4}\sqrt{\pi g_{\textrm{i}}g_{\textrm{r}}g_{\textrm{d}}}+g_{\textrm{d}},
\end{equation}
and $\var\{Z^2\}\triangleq \mathbb{E}\{Z^4\}-(\mathbb{E}\{Z^2\})^2$, which can be obtained from the first four moments of the Gaussian distributed $|h_{\textrm{ir}}|$ and the Rayleigh distributed $|h_{\textrm{d}}|$, and thus is omitted here for brevity.
As a result, the Gamma distribution $\Gamma[k,\theta]$ with the same first and second order moments as $Z^2$ has the shape parameter $k\triangleq (\mathbb{E}\{Z^2\})^2/\var\{Z^2\}$ and scale parameter $\theta\triangleq \var\{Z^2\}/\mathbb{E}\{Z^2\}$.
In the following, we adopt $Z^2 \stackrel{\textrm{approx.}}{\sim} \Gamma[k,\theta]$.



\subsection{IRS Coverage Range}\label{SectionD}
In practice, IRS typically serves UEs in its proximity. Due to this and by exploiting the channel hardening effect of the IRS-reflected path, we propose a practical rule to determine the IRS coverage/serving range $D$.
Specifically, based on the average channel power gains $g_{\textrm{ir}}$ and $g_{\textrm{d}}$ which can be estimated in practice, we define their ratio $\kappa$ based on \eqref{gau} and \eqref{gir} as
{\small
\begin{equation}\label{kappa}
\kappa\triangleq\frac{g_{\textrm{ir}}}{g_{\textrm{d}}}=\frac{G_\textrm{bf} g_{\textrm{i}}g_{\textrm{r}}}{g_{\textrm{d}}}=G_\textrm{bf}\beta\bigg(\frac{\big(r^2+(H_\textrm{B}-H_\textrm{I})^2\big)(d^2+H_\textrm{I}^2)}{l^2+H_\textrm{B}^2}\bigg)^{-\alpha/2},
\end{equation}
}%
which indicates the relative power gain of the IRS-reflected path as compared to the BS-UE direct path.


\rev{Since IRS typically serves nearby UEs, we assume that the distances from the BS to UE 0 and its serving IRS are approximately equal, i.e., $l\approx r$.
As a result, we have $g_{\textrm{d}}\approx g_{\textrm{i}}$ and hence
\begin{equation}\label{kappaApprox}
\kappa\approx G_\textrm{bf} g_{\textrm{r}}=G_\textrm{bf}\beta(d^2+H_\textrm{I}^2)^{-\alpha/2},
\end{equation}
which decreases with $d$, thus suggesting IRS to be deployed as close to its targeted UE(s) as possible.
Therefore, given a certain threshold $\bar \kappa$, we can obtain an estimated IRS coverage range $D$ based on \eqref{kappaApprox} such that $\kappa\geq \bar \kappa$ for $d\leq D$,} which is given by
\begin{equation}\label{D}
D\approx \sqrt{\big(\frac{G_\textrm{bf}\beta}{\bar\kappa}\big)^{2/\alpha}-H_\textrm{I}^2},
\end{equation}
which scales with $O(G_\textrm{bf}^{1/\alpha})$ and thus $O(N^{2/\alpha})$ as $N$ increases.

\subsection{Spatial Throughput}\label{SectionErgodic}
For fixed UE and IRS locations, the distances $l$, $r$ and $d$ are given. For the case with $d> D$, the UE is served by the BS only, with the achievable rate given by
\begin{align}\label{directC}
C&\triangleq \mathbb{E}\{\log_2( 1+ \gamma)\}=\mathbb{E}_{|h_{\textrm{d}}|^2}\{\log_2( 1+ |h_{\textrm{d}}|^2\gamma_0)\}\notag\\
&\stackrel{(a)}{=}E_1\big(\frac{1}{\gamma_0 g_{\textrm{d}}}\big)\exp\big\{\frac{1}{\gamma_0 g_{\textrm{d}}}\big\}\log_2 e \triangleq C_\textrm{d},
\end{align}
where $E_1(\cdot)$ is the exponential integral function available in MATLAB, and $(a)$ is due to the exponentially distributed $|h_{\textrm{d}}|^2$.
On the other hand, for the case with $d\leq D$, we have
\begin{align}
C&\triangleq \mathbb{E}\{\log_2( 1+ \gamma)\}=\mathbb{E}_{Z^2\gamma_0}\{\log_2( 1+ Z^2\gamma_0)\}\notag\\
&\stackrel{(b)}{\approx} {}_{3}F_{1}(k+1,1,1;2;-\theta\gamma_0)k\theta\gamma_0\log_2 e\triangleq C_{\textrm{ir}},\label{Cir}
\end{align}
where $(b)$ is due to that $Z^2\gamma_0$ is approximated by the scaled Gamma distribution $\Gamma[k,\theta\gamma_0]$, 
and the formulae \cite[eq.(4), eq.(8)]{DFrelayCapacity}, 
with ${}_{p}F_{q}$ denoting the generalized hypergeometric function which is available in MATLAB.

Based on the closed-form achievable rates in \eqref{directC} and \eqref{Cir}, the spatial throughput $\bar C$ over the spatial distributions of all UE and IRS locations can be obtained as follows.
Divide the UEs into three disjoint groups in the range $l\in[0,R_1]$, $(R_1,R_2]$ and $(R_2,R_\textrm{c}]$, respectively,
with $R_1\triangleq \max\{0, R_\textrm{in}-\eta D\}$ and $R_2\triangleq \min\{R_\textrm{c}, R_\textrm{out}+\eta D\}$ where $\eta\in(0,1]$ is a given constant to compensate for the boundary effect.
$\bar C$ is then given by
\begin{equation}\label{IntegralC}
\bar C\triangleq \mathbb{E}\{C\}= \frac{S_1}{S}\bar C_1 +\frac{S_2}{S}\bar C_2+\frac{S_3}{S}\bar C_3,
\end{equation}
where $\bar C_1\triangleq \mathbb{E}\{C\}\big|_{l=0}^{R_1}$, $\bar C_2\triangleq \mathbb{E}\{C\}\big|_{l=R_1}^{R_2}$ and $\bar C_3\triangleq \mathbb{E}\{C\}\big|_{l=R_2}^{R_\textrm{c}}$ denote the spatial throughput in the corresponding distance range, respectively, which are weighted by their areas (thanks to uniform UE distribution) given by
\begin{equation}
S_1\triangleq\pi R_1^2, S_3\triangleq\pi(R_\textrm{c}^2-R_2^2), S_2=S-S_1-S_3.
\end{equation}

For the UEs in the range $l\in[0,R_1]$, the pdf of $l$ is given by $f_l(l)\triangleq \frac{2\pi l}{S_1}$. Similarly, for the UEs in the range $l\in(R_2,R_\textrm{c}]$, we have $f_l(l)\triangleq \frac{2\pi l}{S_3}$.
Note that the UEs in these two ranges are mainly served by the BS only, whose achievable rate is given by $C_\textrm{d}$.
Therefore, we have 
\begin{equation}\label{barCd}
\frac{S_1}{S}\bar C_1 +\frac{S_3}{S}\bar C_3\approx \frac{2\pi}{S}\int_{l=0}^{R_1}C_\textrm{d} l\diff l + \frac{2\pi}{S}\int_{l=R_2}^{R_\textrm{c}}C_\textrm{d} l\diff l.
\end{equation}

On the other hand, in the range $l\in(R_1,R_2]$,
we have $\bar C_{2}\triangleq \bar C_{2,1}+\bar C_{2,2}$, where $\bar C_{2,1}$ and $\bar C_{2,2}$ denote the spatial throughput of UEs served by the BS only or jointly served by the IRS, respectively. Specifically, $\bar C_{2,1}$ is given by
\begin{equation}\small\label{C21}
\bar C_{2,1}\triangleq \mathbb{P}\big\{d> D\big\}\mathbb{E}\big\{C\big\}\big|_{l=R_1}^{R_2}= \frac{2\pi}{S_2}e^{-\lambda_\textrm{I}\pi D^2}\int_{l=R_1}^{R_2} C_\textrm{d} l  \diff l,
\end{equation}
where $\mathbb{P}\big\{d> D\big\}\triangleq e^{-\lambda_\textrm{I}\pi D^2}$ with $\lambda_\textrm{I}\approx \frac{M}{\pi(R_\textrm{out}^2-R_\textrm{in}^2)}$ denoting the IRS density, and $f_l(l)\approx \frac{2\pi l}{S_2}$ denotes the approximated pdf of $l$ by treating these UEs to be uniformly distributed.
On the other hand,
$\bar C_{2,2}$ is given by
\begin{align}\label{C22}
\bar C_{2,2}&\triangleq \mathbb{E}\big\{C_{\textrm{ir}}\big\}\big|_{d=0}^{D}\big|_{l=R_1}^{R_2}= \int_{l=R_1}^{R_2}\int_{d=0}^D C_{\textrm{ir}} f_d(d) f_l(l) \diff d\diff l\notag\\
&\stackrel{(c)}{\approx}\int_{r=R_\textrm{in}}^{R_\textrm{out}}\int_{d=0}^D C_{\textrm{ir}}\big|_{l\approx r} f_d(d) f_r(r) \diff d\diff r,
\end{align}
where $(c)$ follows from the approximation that $l\approx r$; $f_r(r)\triangleq \frac{2r}{R_\textrm{out}^2-R_\textrm{in}^2}$ denotes the pdf of $r$ which follows from the uniform distribution of IRS locations in the considered range; and 
$f_d(d)\triangleq e^{-\lambda_\textrm{I}\pi d^2}2\pi\lambda_\textrm{I} d$ denotes the pdf of the nearest UE-IRS distance $d$.
Finally, note that the integration in \eqref{barCd}, \eqref{C21} and \eqref{C22} involves one or two integrals over a finite range with closed-form integrands, respectively, and thus the spatial throughput $\bar C$ in \eqref{IntegralC} can be numerically evaluated efficiently.

\subsection{Benchmark System with Active Relays}

\rev{Consider a benchmark system without using IRS, but instead using the ideal active FD and DF single-antenna relays assuming perfect SIC to provide a performance upper bound.}
For fair comparison with IRS, assume that both the BS and the relay transmit with power $P_0/2$, thus consuming the same total power $P_0$ for each UE.\footnote{\rev{Other setups with different fixed transmit powers $a P_0$ and $(1-a) P_0$ at the BS and relay can be considered, where $0<a<1$ and similar rate performance can be obtained with some changes to the optimal relay deployment range.}}
The achievable rate for the repetition-coded DF relaying \cite{TseRelay} is thus given by 
\begin{equation}\small\label{CkDF}
C\triangleq\mathbb{E}\bigg\{ \min\big\{\log_2\big( 1+\frac{\gamma_0}{2}|h_1|^2\big), \log_2\big( 1+\frac{\gamma_0}{2}|h_{\textrm{d}}|^2+\frac{\gamma_0}{2}|h_2|^2\big)\big\}\bigg\},
\end{equation}
where $|h_1|^2$ and $|h_2|^2$ denote the BS-relay and relay-UE channel power gains, which follow independent exponential distributions with mean $g_1\triangleq \beta \big(u^2+(H_\textrm{B}-H_\textrm{R})^2\big)^{-\alpha/2}$ and $g_2\triangleq \beta \big(v^2+H_\textrm{R}^2\big)^{-\alpha/2}$, respectively, with $H_\textrm{R}$ denoting the relay height, and $u$ and $v$ denoting the horizontal distances of the BS-relay and relay-UE links, respectively.
Since $C$ in \eqref{CkDF} does not admit a closed-form expression, we resort to Monte Carlo (MC) simulations to evaluate it numerically.
On the other hand, each UE can choose to be served by the BS directly, for which the achievable rate is then given by \eqref{directC}.

Finally, consider $M_\textrm{R}$ available relays which are uniformly and randomly deployed in the range $u\in[R_\textrm{in,R},R_\textrm{out,R}]$.
\rev{For practical implementation, assume that 
each UE chooses the optimal mode (direct or relay-aided) and serving relay based on the average channel gains $g_\textrm{d}$, $g_1$ and $g_2$, by substituting them into \eqref{directC} and \eqref{CkDF} and comparing the estimated rates.}
The spatial throughput of the relay-aided system is then evaluated via MC simulatioins.

\section{Numerical Results}\label{SectionNumerical}

\rev{In this section, we verify our analytical results for the spatial throughput $\bar C$ in \eqref{IntegralC} by MC simulations, and further investigate the optimal IRS deployment for maximizing the spatial throughput.}
Each MC simulation result is obtained by averaging over 1000 randomly generated topologies, with 100 fading realizations per channel.
The following parameters are used if not mentioned otherwise: $R_\textrm{c}=250$ m,
$H_\textrm{B}=20$ m, $H_\textrm{I}=1$ m, $H_\textrm{R}=1$ m, $P_0=1$ mW per RB,  $\sigma^2=-110$ dBm, $f_c=2$ GHz, $\alpha=2.5$, $K=500$, $\bar\kappa=0.1$ and $\eta=0.1$.

\begin{figure}
\centering
   \includegraphics[width=1\linewidth,  trim=50 0 50 0,clip]{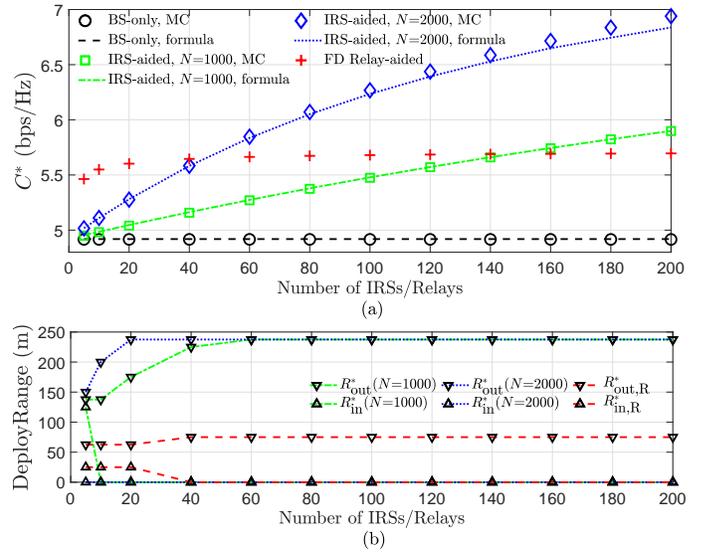}
\caption{\rev{(a) Maximum spatial throughput and (b) optimal deployment range with different number of IRSs/relays.}\vspace{-2ex}}\label{SpatialThroughput}
\end{figure}

First, given the number of IRSs $M$ or relays $M_\textrm{R}$, we obtain their optimal deployment range by 2-dimensional (2D) line search to achieve the maximum spatial throughput in the IRS-aided and relay-aided systems, respectively, where the results are plotted in Fig. \ref{SpatialThroughput}.
It is shown in Fig. \ref{SpatialThroughput}(a) that the analytical results for the IRS-aided system match well with those based on MC simulations.
In addition, the IRS-aided system significantly improves the spatial throughput compared to the baseline BS-only system, and with sufficient IRS elements (e.g., with $N=2000$ and $M\geq 50$) even outperforms the FD relay-aided system with $M_\textrm{R}=M$ active relays.

Fig. \ref{SpatialThroughput}(b) shows the corresponding optimal deployment range 
$[R_\textrm{in,R}^*,R_\textrm{out,R}^*]$ and $[R_\textrm{in}^*,R_\textrm{out}^*]$ for the relay-aided and IRS-aided systems, respectively.
For the relay-aided system with increasing $M_\textrm{R}$, the relays are optimally deployed in the range around $R_\textrm{in,R}^*=0$ and $R_\textrm{out,R}^*=75$ m, while further increasing $M_\textrm{R}$ brings almost no throughput gain.
Moreover, it is observed that on average, a UE using the relay mode at distance $l$ from the BS is served by a relay at distance $u\approx 0.39 l$ from the BS.
\rev{This suggests that the optimal relay position should be around the midpoint between the BS and the served UE, in order to balance between the BS-relay and relay-UE links in \eqref{CkDF}.}

\rev{In contrast, in the IRS-aided system, each IRS provides local coverage for its nearby UEs only, and hence the optimal IRS deployment range expands as $M$ increases}, as shown in Fig. \ref{SpatialThroughput}(b), with more regions covered by the newly deployed IRSs.
\rev{Moreover, as $M$ further increases, the average distance from IRSs to their served UEs gradually decreases, thus providing a sustainable throughput growing.}
On the other hand, the optimal IRS deployment range also depends on the SNR distribution in the cell depending on the BS-UE distance $l$.
Although the IRS provides an approximately constant SNR gain factor $(\kappa+1)$ for a targeted UE at a given distance $d$ away (see \eqref{kappa} and \eqref{kappaApprox}), the achievable rate $C$ in \eqref{Ck} has the largest improvement for UEs located somewhere in the medium SNR region, where the IRSs are more preferably deployed so as to maximize the system spatial throughput.

\begin{figure}
\centering
   \includegraphics[width=1\linewidth,  trim=40 0 70 10,clip]{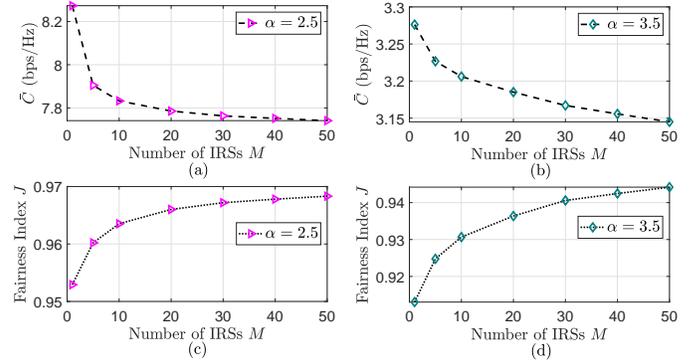}
\caption{\rev{Spatial throughput $\bar C$ and fairness index $J$ under different pathloss exponent $\alpha$ and number of IRSs $M$ with $Q=MN=5000$ and $R=50$ m.}\vspace{-2ex}}\label{MNtradeoff}
\end{figure}

Next, we investigate another important IRS deployment issue, i.e., how many IRSs $M$ should be deployed given a total number of reflecting elements $Q\triangleq MN$?
To investigate the trade-off between $M$ and $N$, we consider an example setup where the IRSs are deployed with equal spacing on a circle with radius $R$ centered at the BS. Consider a group of local UEs in $\mathcal{K}$ located in the ring region $l\in[R-D_\textrm{max},R+D_\textrm{max}]$ which can be potentially served by the IRSs, where $D_\textrm{max}$ is given by \eqref{D} with $N=Q$.
The spatial throughput $\bar C$ and Jain's fairness index\footnote{Jain's fairness index $J$ is defined as $J\triangleq \frac{(\sum_{k\in\mathcal{K}} C_k)^2}{|\mathcal{K}| \sum_{k\in\mathcal{K}} C_k^2}$ with $C_k$ denoting the achievable rate of UE $k$ in \eqref{Ck}, where a higher $J$ represents better fairness.} $J$ for the UEs in $\mathcal{K}$ under different number of IRSs $M$ and pathloss exponent $\alpha$ are shown in Fig. \ref{MNtradeoff}.
It is observed that it is beneficial to assemble more elements into fewer IRSs, in order to maximize the spatial throughput (or UEs' average sum-rate), but at the cost of reduced fairness in the UEs' spatial rate distribution.
This phenomenon is also observed for various other choices of $Q$, $R$ and $\alpha$.
\rev{This is mainly due to the prominent gain factor $\kappa$ given in \eqref{kappaApprox} which scales with $O(N^2)$, thus resulting in significant rate improvement for nearby UEs, whose sum-rate improvement compensates the generally reduced total coverage area with fewer IRSs.}

\section{Conclusions}
This letter investigates a multiuser system aided by multiple IRSs, and characterizes its achievable spatial throughput averaged over channel fading and random IRS/UE locations.
It is shown that the spatial throughput of the IRS-aided system can be superior than that of the FD-relaying based system even assuming the ideal perfect SIC, as the number of IRSs/reflecting elements increases.
It is also shown that the active relays and passive IRSs should be deployed with different strategies in the network to maximize their respective throughputs, while IRSs are preferably deployed to support local rate enhancement and can provide a sustainable throughput growing.
Different from the single-user case where the IRS should be deployed near the BS or user for rate maximization, in the multi-user system, IRSs should be distributed in the whole network to maximize its throughput.
Moreover, an interesting throughput-fairness trade-off is unveiled, where equipping more elements to fewer IRSs helps improve the system throughput, but incurs more spatially varying user rate distribution in the network.

\bibliography{IEEEabrv,BibDIRP}

\newpage

\end{document}